\documentclass[11pt]{article}

\usepackage[margin=1in]{geometry}
\usepackage[T1]{fontenc}
\usepackage[utf8]{inputenc}
\usepackage{lmodern}
\usepackage{microtype}
\usepackage{setspace}

\usepackage{authblk}

\usepackage{amsmath,amssymb}
\usepackage{siunitx}
\usepackage{graphicx}
\usepackage{booktabs}
\usepackage{array}
\usepackage{tabularx}
\usepackage{adjustbox}
\usepackage{multirow}
\usepackage{makecell}
\usepackage{ragged2e}
\usepackage{float}
\usepackage{placeins}
\usepackage{caption}
\usepackage{subcaption}

\newcolumntype{Y}{>{\RaggedRight\arraybackslash}X}
\newcommand{\tablefont}{\small}
\usepackage{enumitem}
\setlist[itemize]{leftmargin=1.5em}
\setlist[enumerate]{leftmargin=1.5em}

\usepackage{xcolor}
\usepackage{xurl}
\usepackage[numbers,sort&compress]{natbib}
\usepackage{hyperref}
\usepackage[nameinlink,noabbrev]{cleveref}

\hypersetup{
    colorlinks=true,
    linkcolor=blue!50!black,
    citecolor=blue!50!black,
    urlcolor=blue!50!black,
    pdftitle={Synthetic Observed-Band Light Curves of Delta Cephei from MESA-RSP Models},
    pdfauthor={Zuhoor Elahi, Christopher Sirola, Wafa Gull}
}

\graphicspath{{./}}

\newcommand{\safeincludegraphics}[2][]{%
  \IfFileExists{#2}{%
    \includegraphics[#1]{#2}%
  }{%
    \fbox{%
      \begin{minipage}[c][0.26\textheight][c]{0.88\linewidth}
      \centering
      \vspace{0.5em}
      \textbf{Missing figure file}\par
      \vspace{0.5em}
      \texttt{\detokenize{#2}}\par
      \vspace{0.5em}
      Place this figure file in the manuscript folder and recompile.
      \end{minipage}}%
  }%
}

\newcommand{\safetableinput}[1]{%
  \IfFileExists{#1}{%
    \input{#1}%
  }{%
    \fbox{%
      \begin{minipage}{0.92\linewidth}
      \centering
      \vspace{0.8em}
      \textbf{Missing table file}\par
      \vspace{0.5em}
      \texttt{\detokenize{#1}}\par
      \vspace{0.8em}
      \end{minipage}}%
  }%
}

\newcommand{\dcep}{\ensuremath{\delta} Cephei}

\newcommand{\Teff}{\ensuremath{T_{\rm eff}}}
\newcommand{\teff}{\Teff}

\newcommand{\logg}{\ensuremath{\log g}}

\newcommand{\feh}{\ensuremath{[{\rm Fe/H}]}}
\newcommand{\MESA}{\textsc{MESA}}
\newcommand{\RSP}{\textsc{RSP}}
\newcommand{\MESARSP}{\MESA-\RSP}

\newcommand{\RSPalfam}{\ifmmode\mathrm{RSP\_alfam}\else\texttt{RSP\_alfam}\fi}
\newcommand{\RSPalfat}{\ifmmode\mathrm{RSP\_alfat}\else\texttt{RSP\_alfat}\fi}
\newcommand{\rspalfam}{\RSPalfam}
\newcommand{\rspalfat}{\RSPalfat}

\title{Synthetic Observed-Band Light Curves of Delta Cephei from MESA-RSP Models}

\author[1,2,*]{Zuhoor Elahi}
\author[1]{Christopher Sirola}
\author[1]{Wafa Gull}

\affil[1]{Department of Physics and Astronomy, University of Southern Mississippi, Hattiesburg, MS, USA}
\affil[2]{Department of Physics, University of Karachi, Karachi, Pakistan}
\affil[*]{Corresponding author: zuhoor.elahi@usm.edu}

\date{}

\begin{document}

\maketitle

% ============================================================
\begin{abstract}
Observed-band light curves provide a stronger test of Cepheid pulsation models than period matching alone, because the measured photometric amplitude depends on the phase-dependent luminosity, temperature, radius, and passband transformation. We construct synthetic observed-band light curves for \dcep{} from nonlinear radial pulsation models computed with MESA-RSP and transform the model outputs \(L(\phi)\), \(\teff(\phi)\), and \(R(\phi)\) into bolometric magnitudes and MIST bolometric-correction magnitudes in Bessell and Gaia passbands. The adopted AAVSO Johnson \(V\)-band template has peak-to-peak amplitude \(\Delta V_{\rm obs}=0.8390~{\rm mag}\). The period-stable reference model with \(\rspalfam=0.60\) gives a MIST-BC \(V\)-band amplitude of only \(\Delta V_{\rm syn}\simeq0.0106~{\rm mag}\). Amplitude-enhanced models increase the synthetic amplitude, with \(\rspalfam=0.425\) giving \(0.0302~{\rm mag}\), \(\rspalfam=0.400\) giving \(0.0344~{\rm mag}\), and the accepted \(\rspalfam=0.400\), \(\rspalfat=0.095\) model giving \(0.0367~{\rm mag}\). The final accepted model therefore reaches only \(4.4\%\) of the observed Johnson \(V\)-band amplitude. These results show that MIST-BC transformations are necessary for a physically meaningful observed-band comparison, but they do not remove the amplitude discrepancy in the present model sequence. The remaining mismatch indicates that the dominant limitation is the small nonlinear pulsation amplitude of the models rather than the bolometric-correction transformation itself.
\end{abstract}


% ============================================================
\section{Introduction}
\label{sec:introduction}
% ============================================================

Classical Cepheids are radially pulsating supergiants whose periods, luminosities, and light-curve shapes provide important constraints on stellar evolution and pulsation theory.  Their use as distance indicators depends on well-calibrated period--luminosity relations \citep{leavitt1912,freedmanmadore1991,freedman2001,riess2022,pietrzynski2019}, but their value as physical laboratories depends on a deeper requirement: stellar models should reproduce the observed temporal behavior of the star, including the pulsation period, luminosity variation, effective-temperature variation, radius variation, and the observed light curve in specific photometric passbands.

The prototype Cepheid \dcep\ is an especially useful benchmark because it is bright, nearby, and extensively observed.  Its visual light curve is strongly asymmetric and has a large Johnson \(V\)-band peak-to-peak amplitude.  In the present work, the adopted AAVSO Johnson \(V\)-band light-curve template \citep{aavso_database} has peak-to-peak amplitude
\begin{equation}
    \Delta V_{\rm obs}=0.8390~{\rm mag}.
    \label{eq:obs_amp}
\end{equation}
This observed amplitude is much larger than the synthetic amplitudes produced by the present MESA-RSP model sequence after transformation into observed bands.

A central lesson from nonlinear pulsation modeling is that period agreement alone is not a sufficient test of a Cepheid model.  The pulsation period is primarily controlled by the global mean density of the star, while the observed-band amplitude depends on the nonlinear hydrodynamic response, the treatment of convection, the thermal structure of the envelope, and the conversion from physical model outputs to observed magnitudes.  Therefore, a model that reproduces the observed period can still fail to reproduce the observed amplitude.

The MESA Radial Stellar Pulsation module (MESA-RSP; MESA denotes Modules for Experiments in Stellar Astrophysics) provides a framework for nonlinear radial stellar pulsation calculations and can output phase-dependent quantities such as luminosity, effective temperature, radius, and pulsation diagnostics \citep{paxton2011mesa,paxton2013mesa,paxton2015mesa,paxton2018mesa,paxton2019mesa}.  However, these physical quantities are not directly equivalent to observed Johnson, Bessell, or Gaia magnitudes.  A bolometric light curve constructed from \(L(\phi)\) is useful, but it is not an observed-band light curve.  Similarly, a blackbody approximation can provide an intermediate \(V\)-band diagnostic, but it does not include realistic stellar-atmosphere effects such as line blanketing and passband-dependent bolometric corrections.  A physically meaningful comparison with observed photometry requires a transformation from \(\{L,\teff,R,\logg,\feh\}\) to magnitudes in real filters.

In this paper, we apply MESA Isochrones and Stellar Tracks bolometric corrections (MIST-BC) to MESA-RSP \dcep\ models in order to construct synthetic multi-band light curves \citep{choi2016mist,dotter2016mist}.  The analysis is restricted to the observed-band transformation and amplitude comparison; detailed RSP parameter calibration and Fourier morphology diagnostics are treated only as context.  The central question is whether physically transformed MESA-RSP outputs reproduce the observed Johnson \(V\)-band amplitude of \dcep.

The transformed light curves show that the MIST-BC transformations improve the physical interpretation of the model--observation comparison, but they do not remove the amplitude discrepancy.  The final accepted model in this study produces a MIST-BC \(V\)-band amplitude of only
\begin{equation}
    \Delta V_{\rm syn}=0.0367~{\rm mag},
\end{equation}
which is
\begin{equation}
    \frac{\Delta V_{\rm syn}}{\Delta V_{\rm obs}}
    =
    \frac{0.0367}{0.8390}
    =
    0.0437
    \approx 4.4\%
\end{equation}
of the observed amplitude.  Equivalently, the observed amplitude is larger than the synthetic amplitude by a factor of approximately \(22.9\).

The structure of the paper is as follows.  Section~\ref{sec:data_models} describes the MESA-RSP models and the observed AAVSO Johnson \(V\)-band comparison data.  Section~\ref{sec:method} presents the synthetic-photometry procedure, including bolometric magnitudes, a blackbody \(V\)-band diagnostic, and the MIST bolometric-correction magnitudes used for the final observed-band amplitudes.  Section~\ref{sec:results} presents the synthetic light-curve amplitudes and the multi-band comparison.  Section~\ref{sec:discussion} discusses the physical meaning of the amplitude mismatch and the limitations of the present approach.  Section~\ref{sec:conclusions} summarizes the main conclusions.

% ============================================================
\subsection*{Scope and relationship to previous work}
The construction of theoretical Cepheid light curves and the transformation of stellar parameters into observational passbands are established topics in the literature \citep[e.g.,][]{bessell1990,bessellmurphy2012,jordi2010,gaia2016,bhardwaj2017,kurbah2023,hocde2024,deka2025}. MESA Isochrones and Stellar Tracks (MIST) bolometric corrections and isochrones provide a widely used framework for mapping stellar parameters into photometric systems \citep{choi2016mist,dotter2016mist}. The purpose of the present work is to apply a consistent bolometric and MIST-bolometric-correction pipeline to the current \dcep{} MESA-RSP model sequence and to quantify the remaining difference between the accepted models and the observed Johnson \(V\)-band amplitude. A complementary semi-empirical reconstruction using photometry, radial velocities, and temperature constraints is presented in \citep{ElahiGull2026SemiEmpirical}; the present paper instead applies the MESA-RSP model outputs directly to the synthetic-photometry transformation.  Detailed harmonic-shape diagnostics are treated in the companion morphology analysis.

\section{Model and Observational Data}
\label{sec:data_models}
% ============================================================

\subsection{MESA-RSP Model Inputs}
\label{subsec:mesa_rsp_inputs}

The synthetic light curves in this paper are constructed from nonlinear radial pulsation models computed with MESA-RSP.  The relevant model outputs are the phase-dependent luminosity, effective temperature, and radius,
\begin{equation}
    L(\phi), \qquad \teff(\phi), \qquad R(\phi),
\end{equation}
where \(\phi\) is pulsation phase.  These quantities provide the physical basis for the bolometric and observed-band transformations used below.

The model sequence comes from the period-calibrated MESA-RSP calculations developed for \dcep.  The sequence produced models with periods close to the observed period and explored changes in RSP control parameters that affect nonlinear amplitude behavior.  The present paper uses these models as inputs to the synthetic-photometry pipeline and emphasizes the observed-band amplitude comparison rather than the full parameter-calibration procedure.

The main models considered are listed in Table~\ref{tab:model_set}.  The MESA-RSP amplitude-control parameter \(\rspalfam\) is used to label the model sequence, and \(\rspalfat\) denotes the time-dependent-convection parameter varied in the final accepted model. The \(\rspalfam=0.60\) model is treated as a period-stable reference model.  The \(\rspalfam=0.425\) and \(\rspalfam=0.400\) models represent amplitude-enhanced cases from the model sequence.  The final accepted model uses
\begin{equation}
    \rspalfam=0.400, \qquad \rspalfat=0.095.
\end{equation}
This final model gives the largest accepted MIST-BC \(V\)-band amplitude in the set considered here, but still remains far below the observed amplitude.

\begin{table}[htbp]
\centering
\caption{MESA-RSP model set used for synthetic observed-band photometry.  The table emphasizes the role of each model in the photometric comparison rather than the detailed RSP calibration history.}
\label{tab:model_set}
\tablefont
\begin{tabularx}{\textwidth}{YccY}
\toprule
Model label & \(\rspalfam\) & \(\rspalfat\) & Role in this paper \\
\midrule
Reference model & 0.60  & baseline/default & Period-stable reference case \\
Amplitude-enhanced model A & 0.425 & baseline/default & First amplitude-enhanced comparison \\
Amplitude-enhanced model B & 0.400 & baseline/default & Stronger amplitude-enhanced comparison \\
Final accepted model & 0.400 & 0.095 & Final synthetic-photometry comparison model \\
\bottomrule
\end{tabularx}
\end{table}

\subsection[Observed Johnson V-Band Light Curve]{Observed Johnson \(V\)-Band Light Curve}
\label{subsec:observed_data}

The observational benchmark is an American Association of Variable Star Observers (AAVSO) Johnson \(V\)-band light curve of \dcep.  The adopted amplitude is taken from the reproducible AAVSO Johnson-\(V\) Fourier-template analysis of \dcep{} \citep{ElahiGull2026AAVSOFourier}.  The data are phase-folded using the adopted Cepheid period and then used to construct an observed \(V\)-band template.  The measured peak-to-peak amplitude of the observed template is
\begin{equation}
    \Delta V_{\rm obs}=0.8390~{\rm mag}.
    \label{eq:observed_delta_v}
\end{equation}
This value is used as the primary amplitude reference throughout the paper.

The primary observational quantity used here is the \(V\)-band peak-to-peak amplitude.  The observed template is used here as an amplitude benchmark for synthetic passband transformations; detailed harmonic-shape diagnostics are treated in the companion morphology analysis.

\subsection{Photometric Bands}
\label{subsec:photometric_bands}

Two photometric products are used for the final amplitude comparisons:

\begin{enumerate}[label=(\roman*)]
    \item a bolometric magnitude diagnostic derived directly from \(L(\phi)\);
    \item MIST bolometric-correction magnitudes in selected Bessell and Gaia passbands.
\end{enumerate}

A blackbody-based \(V\)-band diagnostic is described below as a diagnostic bridge between bolometric and bolometric-correction magnitudes, but the final amplitudes quoted in this paper are the MIST-BC amplitudes.  The MIST-BC transformations are used to compute synthetic light curves in Bessell \(V\), Bessell \(I\), Gaia \(G\), Gaia \(G_{\rm BP}\), and Gaia \(G_{\rm RP}\), subject to the available bolometric-correction table coverage.  The Bessell \(V\) result is the most direct synthetic comparison to the observed AAVSO Johnson \(V\)-band amplitude.

% ============================================================
\section{Synthetic Photometry Method}
\label{sec:method}
% ============================================================

\subsection{Pulsation Phase}
\label{subsec:phase}

The model and observed light curves are compared as functions of pulsation phase.  For a time coordinate \(t\), the phase is defined as
\begin{equation}
    \phi =
    \left(
    \frac{t-t_0}{P}
    \right)
    \bmod 1,
    \label{eq:phase}
\end{equation}
where \(P\) is the adopted pulsation period, \(t_0\) is a reference epoch, and \(\bmod\,1\) denotes taking the fractional part of the elapsed number of pulsation cycles so that \(0 \leq \phi < 1\).  For model outputs, \(\phi\) may be taken directly from the MESA-RSP phase variable when available, or reconstructed from the model time coordinate and period.  For observational data, phase folding places all observations on a common pulsation cycle.

\subsection{Bolometric Magnitude Diagnostic}
\label{subsec:bolometric_diagnostic}

The simplest synthetic magnitude is the bolometric magnitude derived from the phase-dependent luminosity,
\begin{equation}
    M_{\rm bol}(\phi)
    =
    M_{{\rm bol},\odot}
    -
    2.5\log_{10}
    \left[
    \frac{L(\phi)}{L_{\odot}}
    \right].
    \label{eq:mbol}
\end{equation}
The zero point \(M_{{\rm bol},\odot}\) affects the absolute magnitude scale but does not affect the peak-to-peak amplitude.  The bolometric amplitude is therefore
\begin{equation}
    \Delta M_{\rm bol}
    =
    \max_{\phi}\left[M_{\rm bol}(\phi)\right]
    -
    \min_{\phi}\left[M_{\rm bol}(\phi)\right].
    \label{eq:delta_mbol}
\end{equation}

This bolometric magnitude curve is not an observed-band light curve.  It is nevertheless useful because it directly traces the luminosity variation generated by the pulsation model.  If the bolometric amplitude is already very small, then no realistic passband transformation is expected to produce an observed-band amplitude comparable to the measured Johnson \(V\) amplitude unless the temperature-dependent bolometric correction varies strongly over the pulsation cycle.

\subsection[Blackbody V-Band Diagnostic]{Blackbody \(V\)-Band Diagnostic}
\label{subsec:blackbody_diagnostic}

As an intermediate step between a purely bolometric light curve and a full atmosphere-based bolometric-correction transformation, a blackbody \(V\)-band diagnostic can be computed from \(R(\phi)\) and \(\teff(\phi)\).  The phase-dependent blackbody flux through a \(V\)-band response function \(S_V(\lambda)\) may be written schematically as
\begin{equation}
    F_V^{\rm BB}(\phi)
    \propto
    R(\phi)^2
    \int
    B_{\lambda}\!\left[\teff(\phi)\right]
    S_V(\lambda)\,d\lambda,
    \label{eq:bb_flux}
\end{equation}
where \(B_{\lambda}(T)\) is the Planck function.  The corresponding diagnostic magnitude is
\begin{equation}
    m_V^{\rm BB}(\phi)
    =
    -2.5\log_{10}
    \left[
    F_V^{\rm BB}(\phi)
    \right]
    +
    C_V,
    \label{eq:bb_mag}
\end{equation}
where \(C_V\) is an arbitrary zero point.  Because this paper is concerned mainly with amplitudes, the absolute value of \(C_V\) is not important.

The blackbody diagnostic includes the leading effect of temperature-dependent spectral redistribution, but it remains approximate.  Real Cepheid spectra are not blackbodies, and the \(V\)-band flux depends on line blanketing, surface gravity, metallicity, and atmospheric structure.  Therefore, the blackbody diagnostic is used only as a diagnostic bridge between bolometric and MIST-BC magnitudes.

\subsection{MIST Bolometric-Correction Magnitudes}
\label{subsec:mist_bc}

The main synthetic observed-band light curves in this paper are computed using MIST-BC.  For a photometric band \(X\), the absolute magnitude is written as
\begin{equation}
    M_X(\phi)
    =
    M_{\rm bol}(\phi)
    -
    BC_X
    \left[
    \teff(\phi),
    \logg(\phi),
    \feh,
    A_V
    \right],
    \label{eq:mist_bc}
\end{equation}
where \(BC_X\) is the bolometric correction in band \(X\).  The relevant stellar parameters are interpolated phase by phase from the MESA-RSP model outputs.  The surface gravity is computed from
\begin{equation}
    g(\phi)
    =
    \frac{GM}{R(\phi)^2},
    \label{eq:g}
\end{equation}
and therefore
\begin{equation}
    \logg(\phi)
    =
    \log_{10}
    \left[
    \frac{GM}{R(\phi)^2}
    \right],
    \label{eq:logg}
\end{equation}
with \(g\) expressed in cgs units when using standard bolometric-correction tables.

For this study, the most important band is Bessell \(V\), because it is compared directly with the observed AAVSO Johnson \(V\)-band light curve.  Additional bands, including Bessell \(I\), Gaia \(G\), Gaia \(G_{\rm BP}\), and Gaia \(G_{\rm RP}\), are used to evaluate the wavelength dependence of the synthetic amplitude.

\subsection{Amplitude Metrics}
\label{subsec:amplitude_metrics}

For any magnitude curve \(M_X(\phi)\), the peak-to-peak amplitude is defined as
\begin{equation}
    \Delta X
    =
    \max_{\phi}
    \left[
    M_X(\phi)
    \right]
    -
    \min_{\phi}
    \left[
    M_X(\phi)
    \right].
    \label{eq:delta_x}
\end{equation}
Although the magnitude scale is inverted, this definition gives a positive peak-to-peak amplitude.

The synthetic-to-observed amplitude fraction, denoted \(f_{\rm amp}\), is
\begin{equation}
    f_{\rm amp}
    =
    \frac{\Delta V_{\rm syn}}{\Delta V_{\rm obs}},
    \label{eq:f_amp}
\end{equation}
and the amplitude-deficit factor, denoted \(D_{\rm amp}\), is
\begin{equation}
    D_{\rm amp}
    =
    \frac{\Delta V_{\rm obs}}{\Delta V_{\rm syn}}.
    \label{eq:d_amp}
\end{equation}
A perfect amplitude match would have \(f_{\rm amp}=1\) and \(D_{\rm amp}=1\).  Values of \(f_{\rm amp}\ll1\) indicate that the model produces too little photometric variation.

% ============================================================
\section{Results}
\label{sec:results}
% ============================================================

\subsection{Synthetic-Photometry Pipeline}
\label{subsec:pipeline_result}

The analysis begins with the phase-dependent MESA-RSP outputs \(L(\phi)\), \(\teff(\phi)\), and \(R(\phi)\).  These quantities are converted into a bolometric magnitude curve, a blackbody-based \(V\)-band diagnostic, and MIST bolometric-correction light curves in Bessell and Gaia passbands.  This sequence defines the role of the present work: to test whether the transformation from physical pulsation variables into observed-band magnitudes can account for the observed Johnson \(V\)-band amplitude.

\begin{figure}[htbp]
    \centering
    \includegraphics[width=0.95\textwidth,height=0.38\textheight,keepaspectratio]{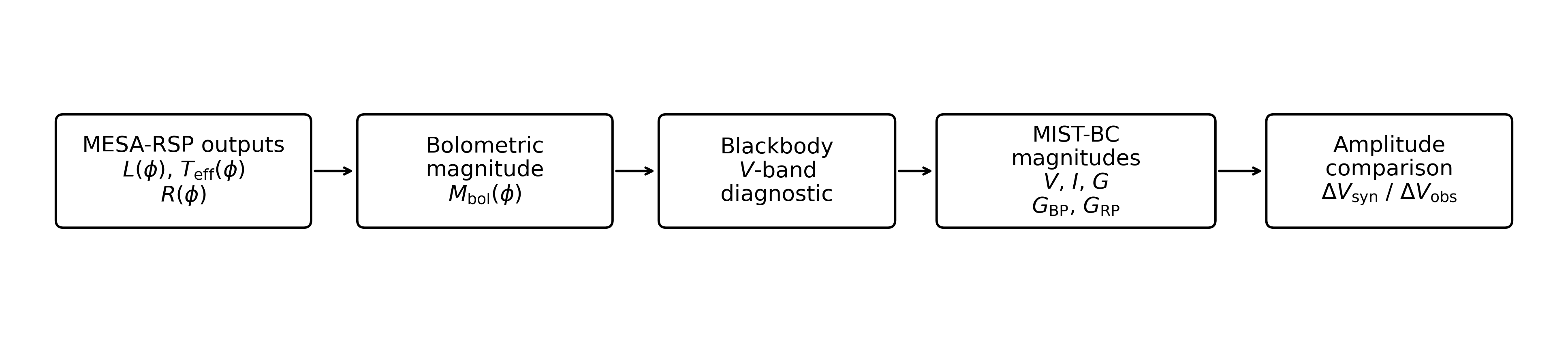}
    \caption{Synthetic-photometry workflow used in this paper.  The MESA-RSP phase-dependent outputs \(L(\phi)\), \(\teff(\phi)\), and \(R(\phi)\) are converted first into \(M_{\rm bol}\), then into a blackbody \(V\)-band diagnostic, and finally into MIST bolometric-correction magnitudes in Bessell \(V\), Bessell \(I\), Gaia \(G\), Gaia \(G_{\rm BP}\), and Gaia \(G_{\rm RP}\).}
    \label{fig:photometry_pipeline}
\end{figure}

\subsection{Physical Phase Curves of the Accepted Model}
\label{subsec:physical_phase_curves}

The passband-dependent synthetic amplitudes are controlled by the underlying luminosity, effective-temperature, and radius variations.  The luminosity curve sets the bolometric magnitude variation, while \(\teff(\phi)\), \(R(\phi)\), and \(\logg(\phi)\) determine how the bolometric flux is redistributed among optical and Gaia passbands through the bolometric-correction tables.

\begin{figure}[htbp]
    \centering
    \includegraphics[width=0.95\textwidth,height=0.42\textheight,keepaspectratio]{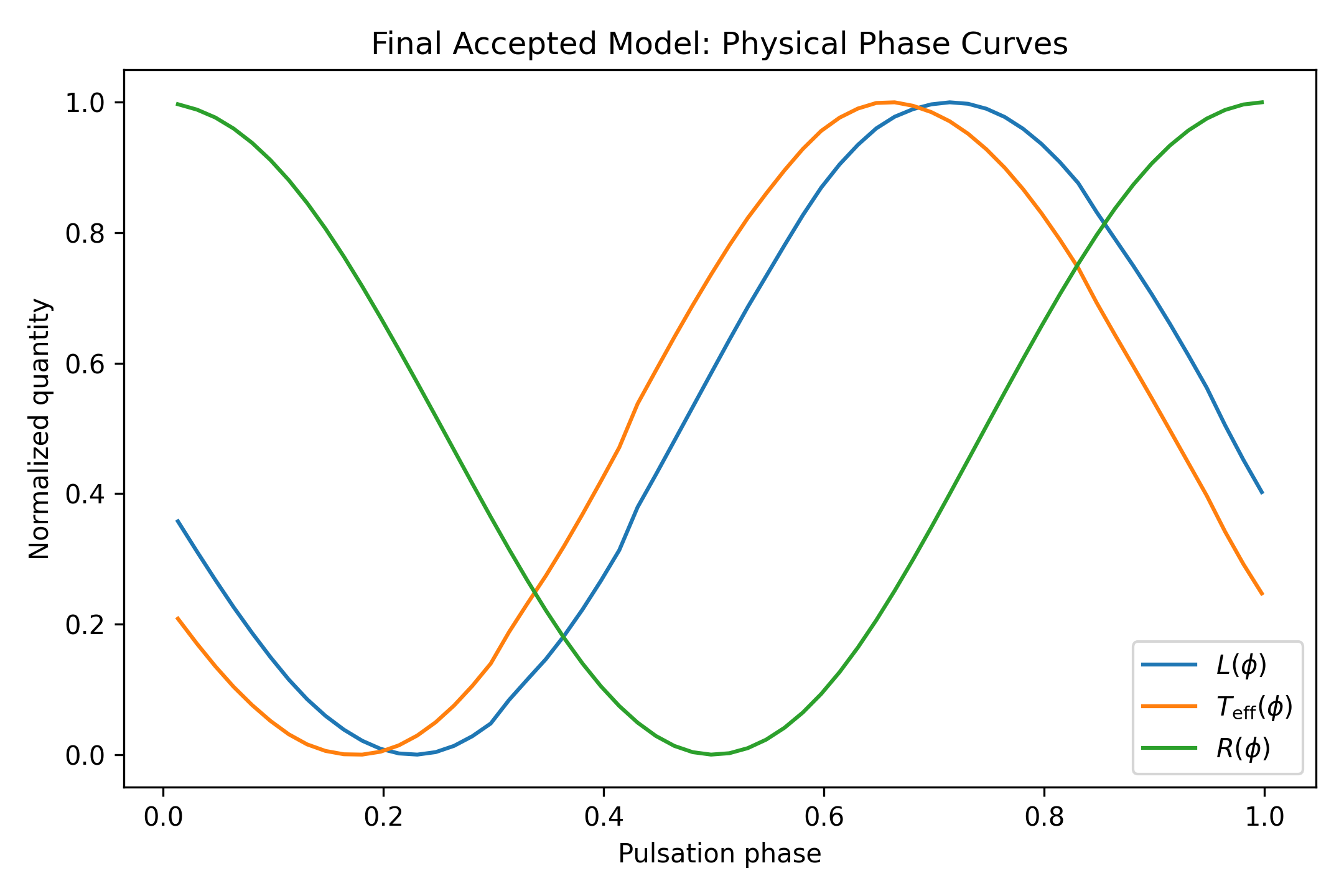}
    \caption{Phase-dependent physical quantities for the final accepted MESA-RSP model: luminosity \(L(\phi)\), effective temperature \(\teff(\phi)\), and radius \(R(\phi)\).  These curves are the direct inputs to the bolometric and MIST bolometric-correction transformations.}
    \label{fig:physical_phase_curves}
\end{figure}

\subsection[Bolometric and MIST-BC V-Band Transformations]{Bolometric and MIST-BC \(V\)-Band Transformations}
\label{subsec:bol_bb_mist_results}

The bolometric diagnostic and MIST-BC \(V\)-band light curve represent two stages of the transformation from MESA-RSP physical output to observed-band photometry.  The bolometric diagnostic isolates the luminosity variation \(L(\phi)\).  The MIST-BC curve then applies the adopted bolometric-correction transformation using \(\teff\), \(\logg\), composition, and extinction assumptions.

This comparison tests whether the observed amplitude deficit is mainly a passband-transformation problem.  If the synthetic amplitude became comparable to the observed amplitude only after the MIST-BC step, then the discrepancy would mainly reflect the difference between bolometric and observed-band photometry.  Instead, the MIST-BC \(V\)-band amplitude remains far below the observed Johnson \(V\)-band amplitude, indicating that the mismatch is not removed by a physically motivated bolometric-correction transformation.

\begin{figure}[htbp]
    \centering
    \includegraphics[width=0.95\textwidth,height=0.40\textheight,keepaspectratio]{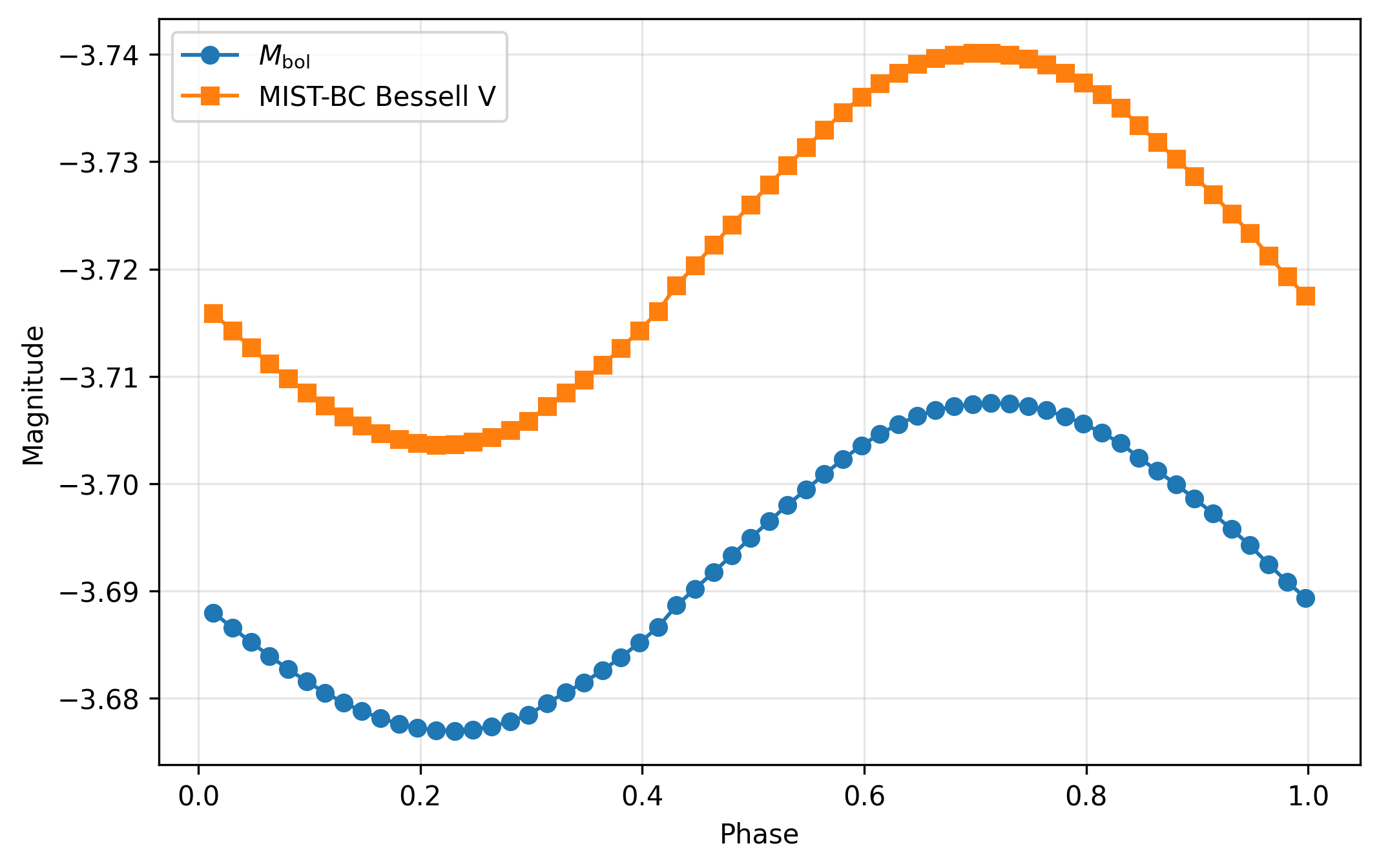}
    \caption{Comparison of the bolometric magnitude curve and MIST-BC Bessell \(V\)-band light curve for the final accepted \(\rspalfam=0.400,\ \rspalfat=0.095\) model.  The figure shows how the amplitude changes as the model output is transformed from bolometric luminosity into an observed-band magnitude.}
    \label{fig:bol_bb_mist_v}
\end{figure}

\subsection[Observed and Synthetic V-Band Amplitude Deficit]{Observed and Synthetic \(V\)-Band Amplitude Deficit}
\label{subsec:v_amplitude_deficit}

The observed AAVSO Johnson \(V\)-band template has peak-to-peak amplitude
\begin{equation}
    \Delta V_{\rm obs}=0.8390~{\rm mag}.
\end{equation}
This amplitude is the primary observational benchmark for the present work.

Table~\ref{tab:v_amplitude_deficit} gives the main \(V\)-band amplitude comparison.  The period-stable reference model with \(\rspalfam=0.60\) produces a MIST-BC \(V\)-band amplitude of approximately \(0.0106~{\rm mag}\), only \(1.26\%\) of the observed amplitude.  The amplitude-enhanced models increase the synthetic \(V\)-band amplitude, but not enough to match the observed light curve.  The \(\rspalfam=0.425\) model gives \(\Delta V_{\rm syn}=0.0302~{\rm mag}\), while the \(\rspalfam=0.400\) model gives \(\Delta V_{\rm syn}=0.0344~{\rm mag}\).

The final accepted model with \(\rspalfam=0.400\) and \(\rspalfat=0.095\) gives
\begin{equation}
    \Delta V_{\rm syn}=0.0367~{\rm mag}.
\end{equation}
This is the largest accepted \(V\)-band amplitude in the model set considered here, but it is still much smaller than the observed value.

\begin{table}[htbp]
\centering
\caption{Observed and synthetic \(V\)-band amplitudes.  The observed benchmark is the AAVSO Johnson \(V\)-band amplitude.  Synthetic values are MIST-BC \(V\)-band amplitudes.}
\label{tab:v_amplitude_deficit}
\tablefont
\begin{tabularx}{\textwidth}{Ycccc}
\toprule
Model & \makecell{\(\Delta V\)\\(mag)} & \(f_{\rm amp}\) & Percentage & \(D_{\rm amp}\) \\
\midrule
Observed AAVSO Johnson \(V\) & 0.8390 & 1.0000 & 100.0\% & 1.0 \\
\(\rspalfam=0.60\) & 0.0106 & 0.0126 & 1.26\% & 79.2 \\
\(\rspalfam=0.425\) & 0.0302 & 0.0360 & 3.60\% & 27.8 \\
\(\rspalfam=0.400\) & 0.0344 & 0.0410 & 4.10\% & 24.4 \\
\(\rspalfam=0.400,\ \rspalfat=0.095\) & 0.0367 & 0.0437 & 4.37\% & 22.9 \\
\bottomrule
\end{tabularx}
\end{table}

The improvement from the reference model to the final accepted model is significant in relative terms:
\begin{equation}
    \frac{0.0367}{0.0106}
    \simeq
    3.46.
\end{equation}
However, this increase is not sufficient in absolute terms.  Even after this improvement, the final accepted model remains low by a factor of about \(22.9\) relative to the observed Johnson \(V\)-band amplitude.

\begin{figure}[htbp]
    \centering
    \includegraphics[width=0.90\textwidth,height=0.38\textheight,keepaspectratio]{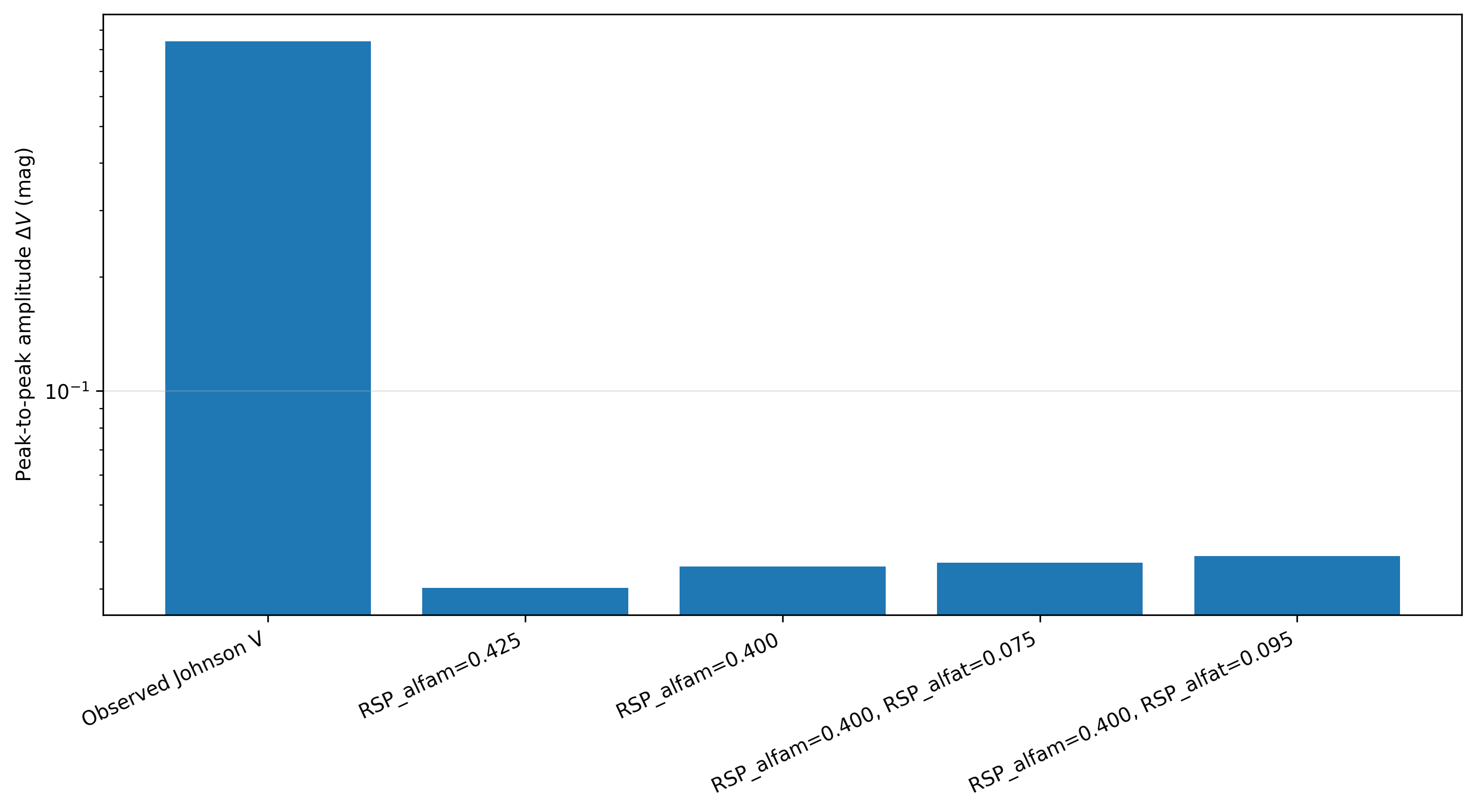}
    \caption{Observed-versus-synthetic \(V\)-band amplitude deficit.  The figure emphasizes absolute peak-to-peak amplitudes and amplitude ratios rather than normalized light-curve morphology.  The final accepted model reaches \(\Delta V_{\rm syn}=0.0367~{\rm mag}\), corresponding to only \(4.4\%\) of the observed \(\Delta V_{\rm obs}=0.8390~{\rm mag}\).}
    \label{fig:v_amplitude_deficit}
\end{figure}

\subsection{Synthetic Multi-Band MIST-BC Amplitudes}
\label{subsec:multiband_results}

The MIST bolometric-correction pipeline also produces synthetic amplitudes in multiple passbands.  Table~\ref{tab:multiband_amplitudes} summarizes the final accepted model amplitudes in Bessell and Gaia filters.  Different filters respond differently to the same underlying pulsation cycle.  Bluer filters are generally more sensitive to temperature variations, while redder filters are less sensitive to temperature and more closely trace changes in radius and luminosity.  Therefore, the band dependence of the synthetic amplitudes helps identify whether the model is limited primarily by the luminosity variation, the temperature variation, or the photometric transformation.

The final accepted-model MIST-BC pipeline gives synthetic amplitudes of 0.0365 mag in Bessell \(V\), 0.0226 mag in Bessell \(I\), 0.0330 mag in Gaia \(G_{\rm EDR3}\), 0.0403 mag in Gaia \(G_{\rm BP,EDR3}\), and 0.0242 mag in Gaia \(G_{\rm RP,EDR3}\).  These values confirm that changing passband changes the synthetic amplitude, but no passband approaches the observed Johnson \(V\)-band amplitude.

\begin{table}[htbp]
\centering
\caption{Multi-band MIST-BC amplitudes for the final accepted model.}
\label{tab:multiband_amplitudes}
\tablefont
\begin{tabularx}{\textwidth}{YccY}
\toprule
Band & \makecell{Synthetic\\amplitude (mag)} & \makecell{Relative to\\observed \(V\)} & Comment \\
\midrule
Bessell \(V\) & 0.0365 & 0.0435 & Final accepted \(V\)-band result \\
Bessell \(I\) & 0.0226 & 0.0270 & Redder Bessell band \\
Gaia \(G_{\rm EDR3}\) & 0.0330 & 0.0393 & Gaia broad band \\
Gaia \(G_{\rm BP,EDR3}\) & 0.0403 & 0.0480 & Largest synthetic amplitude in this set \\
Gaia \(G_{\rm RP,EDR3}\) & 0.0242 & 0.0288 & Gaia red band \\
\bottomrule
\end{tabularx}
\end{table}

\begin{figure}[htbp]
    \centering
    \includegraphics[width=0.90\textwidth,height=0.38\textheight,keepaspectratio]{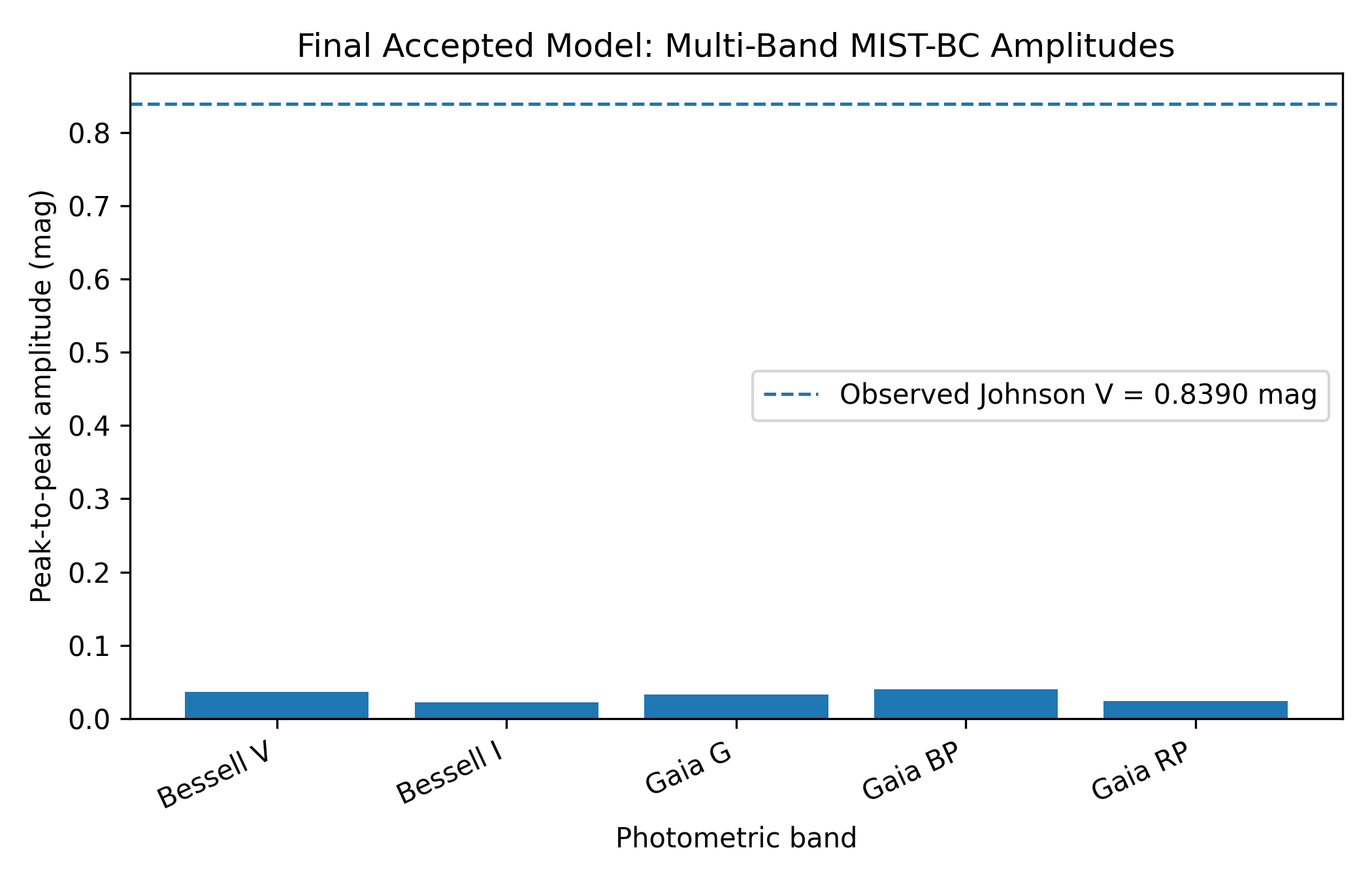}
    \caption{Multi-band MIST-BC synthetic amplitudes for the final accepted model.  The comparison across Bessell \(V\), Bessell \(I\), Gaia \(G\), Gaia \(G_{\rm BP}\), and Gaia \(G_{\rm RP}\) shows that the amplitude deficit persists across the tested passbands.}
    \label{fig:multiband_amplitudes}
\end{figure}

\subsection{Compact Final Amplitude Summary}
\label{subsec:compact_summary}

The final accepted model reaches only a small fraction of the observed Johnson \(V\)-band amplitude.  The compact summary in Table~\ref{tab:final_amplitude_summary} separates the observed constraint, the final synthetic result, and the inferred amplitude deficit factor.

\begin{table}[htbp]
\centering
\caption{Compact amplitude summary for the final accepted model.}
\label{tab:final_amplitude_summary}
\tablefont
\begin{tabularx}{0.85\textwidth}{Yc}
\toprule
Quantity & Value \\
\midrule
Observed Johnson \(V\)-band amplitude, \(\Delta V_{\rm obs}\) & \(0.8390~{\rm mag}\) \\
Final MIST-BC \(V\)-band amplitude, \(\Delta V_{\rm syn}\) & \(0.0367~{\rm mag}\) \\
Synthetic-to-observed amplitude fraction, \(f_{\rm amp}\) & \(0.0437\) \\
Synthetic amplitude as percentage of observed amplitude & \(4.37\%\) \\
Observed-to-synthetic amplitude deficit factor, \(D_{\rm amp}\) & \(22.9\) \\
\bottomrule
\end{tabularx}
\end{table}

% ============================================================
\section{Discussion}
\label{sec:discussion}
% ============================================================

\subsection{Bolometric Corrections Are Necessary}
\label{subsec:bc_necessary}

A direct comparison between MESA-RSP luminosity and observed Johnson \(V\)-band photometry is physically incomplete.  The observed \(V\)-band flux is not determined by bolometric luminosity alone.  It depends on the distribution of flux across wavelength, which changes over the pulsation cycle as \(\teff\), \(R\), \(\logg\), and the atmospheric structure vary.  Therefore, bolometric corrections are necessary for a meaningful observed-band comparison.

This point is especially important for Cepheids because their visual amplitudes are strongly affected by temperature variation.  During the pulsation cycle, the star changes both size and effective temperature.  These variations alter the bolometric luminosity, but they also redistribute radiation across optical and near-infrared bands.  A physically motivated bolometric-correction table accounts for this redistribution more realistically than either a bolometric diagnostic or a blackbody approximation.

The MIST-BC transformation therefore represents an important methodological step.  It converts the theoretical MESA-RSP outputs into magnitudes that can be compared with observed-band light curves.  Without this step, the comparison would risk confusing bolometric variation with passband-specific photometric variation.

\subsection{Bolometric Corrections Do Not Solve the Amplitude Mismatch}
\label{subsec:bc_do_not_solve}

Although MIST-BC transformations are necessary, the results show that they are not sufficient to solve the amplitude problem.  The final accepted model gives
\begin{equation}
    \Delta V_{\rm syn}=0.0367~{\rm mag},
\end{equation}
while the observed AAVSO Johnson \(V\)-band amplitude is
\begin{equation}
    \Delta V_{\rm obs}=0.8390~{\rm mag}.
\end{equation}
Thus,
\begin{equation}
    f_{\rm amp}=0.0437,
\end{equation}
meaning that the model reaches only about \(4.4\%\) of the observed amplitude.

This result has an important interpretation.  The amplitude mismatch is not merely caused by using an inappropriate photometric transformation.  If the RSP model produces too little variation in the underlying physical quantities \(L(\phi)\), \(\teff(\phi)\), and \(R(\phi)\), then a realistic bolometric correction cannot create a large observed-band amplitude from an intrinsically weak pulsation.  The MIST-BC transformation can redistribute the model flux into realistic passbands, but it cannot fully compensate for insufficient hydrodynamic amplitude.

\subsection{Physical Interpretation}
\label{subsec:physical_interpretation}

The small synthetic \(V\)-band amplitudes suggest that the present model sequence underpredicts the nonlinear pulsation amplitude of \dcep.  A related fixed-model opacity-sensitivity experiment shows that native MESA opacity choices measurably affect period matching and amplitude-growth diagnostics, but do not by themselves remove the observed-amplitude discrepancy \citep{ElahiSirolaGull2026Opacity}.  Several physical and numerical factors may contribute:

\begin{enumerate}[label=(\roman*)]
    \item \textbf{Hydrodynamic amplitude:} The limit-cycle amplitude produced by the RSP model may be intrinsically too small for the adopted stellar parameters and convection settings.

    \item \textbf{Convection--pulsation coupling:} The amplitude of a Cepheid model is sensitive to time-dependent convection parameters.  Changes in convective transport and turbulent dissipation can alter the nonlinear amplitude without necessarily destroying the period match.

    \item \textbf{Static atmosphere approximation:} MIST bolometric corrections are applied phase by phase using stellar parameters from the pulsation model.  This is physically useful, but it does not represent a fully dynamic Cepheid atmosphere with velocity fields, shocks, and phase-dependent line formation.

    \item \textbf{Temperature-amplitude limitation:} If the model temperature variation is too small, then the \(V\)-band amplitude will remain too small even after applying realistic bolometric corrections.

    \item \textbf{Parameter degeneracy:} A model can match the period through an appropriate mean density while still failing in observed-band amplitude.  This reinforces the need to evaluate period agreement and passband amplitude as distinct constraints.
\end{enumerate}

The present paper does not attempt to identify a unique correction among these possibilities.  Instead, it establishes the observational-band amplitude deficit after a physically motivated photometric transformation.

\subsection{Relation to Complementary Model Diagnostics}
\label{subsec:relationship_diagnostics}

The observed-band amplitude comparison presented here is one component of a broader validation of the \dcep{} MESA-RSP model sequence.  This paper isolates the synthetic-photometry question: how \(L(\phi)\), \(\teff(\phi)\), and \(R(\phi)\) map into observed passbands and how much amplitude remains after that transformation.  Detailed Fourier morphology diagnostics, including harmonic amplitude ratios, phase combinations, rise fraction, and asymmetry measures, are treated in the companion morphology analysis.

The main result from the present analysis is that MIST bolometric corrections are required for a fair photometric comparison, but they do not by themselves remove the amplitude mismatch.

\subsection{Limitations}
\label{subsec:limitations}

Several limitations should be kept in mind.

First, the synthetic photometry is based on phase-by-phase interpolation in bolometric-correction tables.  This is a standard and useful method, but it is not equivalent to computing a fully dynamic atmosphere model for a pulsating Cepheid.

Second, the main observational comparison in this paper uses Johnson \(V\)-band data.  Additional observed light curves in \(B\), \(I\), Gaia, or near-infrared bands would provide stronger constraints on the wavelength dependence of the amplitude mismatch.

Third, the paper does not perform a harmonic morphology calibration.  The amplitude comparison is necessary, but not sufficient, for validating the model.  Even if a model reproduced the observed amplitude, it would still need to reproduce the detailed shape and harmonic structure of the observed light curve.

Fourth, the multi-band results depend on the adopted bolometric-correction tables, interpolation scheme, metallicity, extinction assumptions, and filter definitions.  These assumptions should be kept fixed and documented when comparing different RSP models.

Finally, the present results apply to the specific MESA-RSP model sequence used here.  They should not be interpreted as a general failure of MESA-RSP or MIST bolometric corrections.  Rather, they show that the current \dcep\ model sequence still underpredicts the observed \(V\)-band amplitude after a physically necessary observed-band transformation.

% ============================================================
\section{Conclusions}
\label{sec:conclusions}
% ============================================================

This paper constructed synthetic observed-band light curves for \dcep\ from MESA-RSP nonlinear radial pulsation models using bolometric magnitudes, a blackbody \(V\)-band diagnostic, and MIST bolometric corrections.  The main conclusions are as follows.

\begin{enumerate}
    \item MESA-RSP outputs \(L(\phi)\), \(\teff(\phi)\), and \(R(\phi)\) must be transformed into observed photometric bands before being compared with real light curves.  Bolometric luminosity alone is not equivalent to Johnson \(V\), Bessell \(I\), or Gaia-band photometry.

    \item The observed AAVSO Johnson \(V\)-band light curve of \dcep\ has peak-to-peak amplitude
    \[
        \Delta V_{\rm obs}=0.8390~{\rm mag}.
    \]

    \item The period-stable reference model with \(\rspalfam=0.60\) gives a MIST-BC \(V\)-band amplitude of only
    \[
        \Delta V_{\rm syn}\simeq0.0106~{\rm mag},
    \]
    corresponding to about \(1.26\%\) of the observed amplitude.

    \item Amplitude-enhanced models increase the synthetic \(V\)-band amplitude.  The \(\rspalfam=0.425\) model gives
    \[
        \Delta V_{\rm syn}=0.0302~{\rm mag},
    \]
    and the \(\rspalfam=0.400\) model gives
    \[
        \Delta V_{\rm syn}=0.0344~{\rm mag}.
    \]

    \item The final accepted model with \(\rspalfam=0.400\) and \(\rspalfat=0.095\) gives
    \[
        \Delta V_{\rm syn}=0.0367~{\rm mag}.
    \]
    This is only
    \[
        \frac{0.0367}{0.8390}=0.0437,
    \]
    or about \(4.4\%\), of the observed Johnson \(V\)-band amplitude.

    \item Therefore, the observed \(V\)-band amplitude is larger than the final synthetic amplitude by a factor of approximately \(22.9\).

    \item MIST bolometric corrections are physically necessary for observed-band comparisons, but they do not solve the amplitude mismatch in the present \dcep\ MESA-RSP model sequence.

    \item Further modeling will combine improved RSP amplitude calibration, additional multi-band observational constraints, and more advanced atmosphere treatments where appropriate.  Detailed Fourier morphology diagnostics are treated separately in the companion morphology analysis.
\end{enumerate}

The central conclusion is that the present models can be transformed into physically meaningful observed-band light curves, but the resulting amplitudes remain much too small.  The amplitude discrepancy is therefore a real modeling problem, not simply a consequence of comparing bolometric model outputs with observed \(V\)-band data.

% ============================================================
\section*{Acknowledgments}
% ============================================================

The authors acknowledge the developers of \MESA{} and the broader stellar pulsation community whose tools and methods made this work possible.

% ============================================================
\section*{Data and Code Availability}
% ============================================================

The calculations were performed with \MESA{} version r24.08.1 using a \dcep{}-specific \MESARSP{} model sequence.  The primary model outputs used in this paper are the phase-dependent luminosity, effective temperature, radius, surface gravity, bolometric magnitude, and MIST bolometric-correction synthetic magnitudes.  The reduced data products include the final accepted phase table, the multi-band MIST-BC amplitude summary, and the observed-versus-synthetic amplitude-comparison tables.  Post-processing scripts were used to construct bolometric magnitude curves, apply MIST bolometric corrections in the selected Bessell and Gaia passbands, compute peak-to-peak amplitudes, and generate the figures and tables presented in this paper.  The inlists, post-processing scripts, and reduced diagnostic tables are available from the corresponding author upon reasonable request.

% ============================================================
\bibliographystyle{unsrtnat}
\bibliography{references}

\end{document}